# Near total reflection X-ray photoelectron spectroscopy: Quantifying chemistry at solid/liquid and solid/solid interfaces

H.P. Martins,[1,2] G. Conti,[1,3] I. Cordova,[4] L. Falling,[1] H. Kersell,[1] F. Salmassi,[4] E. Gullikson,[4] I. Vishik,[3] C. Baeumer,[5] P. Naulleau,[4] C. M. Schneider,[3,6] S. Nemsak[1]

[1] Advanced Light Source, Lawrence Berkeley National Laboratory, Berkeley, CA 94720, USA
[2] Department of Physics, Carnegie Mellon University, Pittsburgh, PA 15213, USA
[3] Department of Physics, University of California, Davis, CA 95616, USA
[4] Center for X-ray optics, Lawrence Berkeley National Laboratory, Berkeley, CA 94720, USA
[5] MESA+ Institute for Nanotechnology, University of Twente, Faculty of Science and Technology, 7500 AE Enschede, Netherlands
[6] Peter-Gruenberg-Institut-6, Forschungszentrum Juelich, Juelich, 52425, Germany

*Abstract*

Near total reflection regime has been widely used in X-ray science, specifically in grazing incidence small angle X-ray scattering and in hard X-ray photoelectron spectroscopy. In this work, we introduce some practical aspects of using near total reflection in ambient pressure X-ray photoelectron spectroscopy and apply this technique to study chemical concentration gradients in a substrate/photoresist system. Experimental data are accompanied by X-ray optical and photoemission simulations to quantitatively probe the photoresist and the interface with the depth accuracy of ~1 nm. Together, our calculations and experiments confirm that near total reflection X-ray photoelectron spectroscopy is a suitable method to extract information from buried interfaces with highest depth-resolution, which can help address open research questions regarding our understanding of concentration profiles, electrical gradients, and charge transfer phenomena at such interfaces. The presented methodology is especially attractive for solid/liquid interface studies, since it provides all the strengths of a Bragg-reflection standing-wave spectroscopy without the need of an artificial multilayer mirror serving as a standing wave generator, thus dramatically simplifying the sample synthesis.

**Introduction**

The interfaces between solids, liquids, and gases are of significant technological and environmental importance. For example, high-value chemicals and sustainable fuels are produced on electrocatalyst surfaces immersed in liquid electrolyte.[1] Similarly related to interfaces, oxidation chemistry in organic aerosols are affecting Earth's climate and air quality. The charge transfer reactions depend on the interface properties, yet almost all our understanding of electrocatalytic activity trends comes from the bulk material properties in the as-prepared state. We still lack interface-sensitive probes of the composition and electronic structure of buried interfaces like the solid/liquid interface.

In an effort to quantify the composition and probe the electronic state of interfaces, X-ray spectroscopy has been adapted and developed to work at pressures different from high or ultrahigh vacuum, the pressure at which it is usually operated. Having reached the necessary technical maturity in the 2000s, ambient pressure X-ray photoelectron spectroscopy, or AP-XPS, is employed more often and sparked a steep rise of publications using the technique.[2,3] But most of these studies have been confined to the solid/gas[4] or liquid/gas[5] interface due to the typically short attenuation length of photoelectrons in solids and liquids. For example, the effective attenuation length for electrons with 100 and 1000 eV kinetic energy is 0.6 and 2.4 nm in $SiO_2$ (based on TPP-2M formula[6]), 2 and 8 nm in liquid water.[7] By contrast, buried solid/liquid or solid/solid interfaces are still considered a challenge for quantitative XPS, because photoemission often yields only a small fraction of the usable information from the interface. Increasing the information depth, e.g. by using higher excitation energies in hard X-ray photoemission, only allows experimentalists to look deeper,[8] without interface specificity.[9] One approach to resolve buried interfaces in XPS is to employ Bragg reflection X-ray standing waves (SWs), a method explored for solid/solid interfaces,[10] which was recently also applied to solid/liquid interfaces.[11,12] These SWs are periodic fields of antinodes and nodes created by interference of incoming and reflected photons. Using multilayer mirrors as SW generators, it has been demonstrated that the signal originating from the antinodes is strongly amplified and that the antinodes could be shifted perpendicular to the surface with a sub-nm precision.[13,14] A prerequisite for achieving such precision is, however, the accurate preparation of multilayer mirrors and detailed knowledge about layer thicknesses and their optical properties, which is often not feasible.

In this work, we propose another approach to achieve comparable depth resolution in XPS analysis of solid/liquid interfaces without the need for multilayer mirrors. The methodology we suggest herein makes use of X-ray optical effects in the near total reflection (NTR) regime. Such effects are widely applied in grazing incidence X-ray scattering and other photon-in photon-out experiments, as it can selectively tune the penetration depth of X-rays by changing its energy and incidence angle. The photoelectron spectroscopy in NTR has been previously successfully applied to solid/solid interfaces,[15,16] and we discuss its extension to solid/liquid interfaces in this work. In a first step, we perform X-ray optical calculations to demonstrate how interface-sensitive information can be extracted from XPS of a solid/liquid/gas system in the NTR regime, representative of typical operando XPS experiments for the solid/liquid interface.[17,18] Next, we demonstrate experimentally that the suggested approach indeed yields the desired interface specificity. For this purpose, we use a model sample in a "glassy-like phase", i.e. an organic photoresist for extreme ultraviolet (EUV) lithography on silica, representing the simulated solid/liquid interface appropriately.

EUV lithography is the leading-edge technology for sub-nm patterning and miniaturization. The most popular class of EUV photoresists are the chemically amplified resists (CARs).[19] The standard CARs consist of a polymer resin, a photo-acid generator which provides sensitivity to ultraviolet light, and a dissolution inhibitor which provides solubility before and after EUV exposure. After spin-on deposition of a thin layer and a post-application bake, the resist is in a disordered state that can represent the liquid side of a solid/liquid interface.

Our experiments reveal the composition of this layer with high depth resolution and confirm the expectations from the X-ray optical calculations. Together, our calculations and experiments confirm that NTR-XPS is indeed a suitable method to extract information from solid/liquid interfaces with highest depth-resolution, which can help address open research questions regarding our understanding and the exploitation of charge transfer phenomena at such interfaces.

**Methods**

Photoemission experiments were performed using the HAXPES endstation at the beamline 9.3.1 of the Advanced Light Source. Measurements were done using photon energy of 3 keV, resulting in energy resolution better than 500 meV. Energy calibration was done using Si 1s core-level, aligning peaks for elemental and oxidized species to 1839.2[20] and 1843.8 eV,[21] respectively. The angle between the hemispherical electron analyzer and the beamline was set to 90°, the incidence angle of X-rays was controlled by polar rotation of the sample. Samples were rotated with respect to the incident beam in grazing incidence in 0.02° steps. Photoemission spectra were analyzed and fitted by KolXPD software package,[22] using Shirley background subtraction and symmetrical Voigt-shape function to model individual peaks. The evolution of the core-level peak area as a function of incident angle, otherwise called a rocking-curve (RC), is then used for a comparison between experimental and simulated data, as described below.

For NTR experiments, the sample is illuminated at low incidence angles and the angle is varied in precise steps to tune the penetration depth: At an incidence angle below the critical angle of near-total external reflection, $\theta_c$, the sensitivity towards surface structures is dramatically enhanced, which has widely been made use of in X-ray scattering experiments,[23,24,25] even in coherent regimes.[26] Incidence angles above $\theta_c$, enhance sensitivity to subsurface structures. At the same time, due to the interference of incident and reflected waves, a strong X-ray standing wave with a relatively large period is formed and creates a periodic electric field modulation in and above the sample.

Our X-ray optical simulations predict these X-ray optical effects and the resulting depth-dependent X-ray intensities as a function of incidence angle. The simulations were performed using the X-ray optical and photoemission simulation package called YXRO.[27] X-ray optical constants were obtained from the CXRO database published in ref. 28. X-ray simulations involve solving Snell's equation for the solid/solid and solid/liquid interfaces in small depth increments. The photoemission yield is subsequently calculated using differential photoionization cross-sections[29] and by integrating the exponentially attenuated signal using the inelastic mean free path as predicted by Tanuma-Powell-Penn formula.[6]

For the analysis and fitting of the experimental data, a global optimization algorithm (called SWOPT as "Standing Wave Optimizer") based on a surrogate model and a black box optimizer was used. A quality of the fit between the simulation and experimental data is judged by the error function, which is defined as a sum of squared residuals for all RCs. Further details on this SWOPT optimization algorithm can be found in the ref. 30.

CARs are typically used based on the process shown in Figure 2a: After deposition by spin-on coating on a chosen substrate, they undergo a) post-application bake (PAB), which removes the solvent and the water used during the spin-on deposition; b) exposure to EUV light; c) post exposure bake (PEB); d) development. Commercial CAR films with a nominal density of 1.4 g/cm$^3$ were spin-coated[31] onto substrates at 4000 rpm and yielded an initial thickness of 14 nm, as measured by ellipsometry. The accuracy of the ellipsometry measurements depends on the correct determination of the material density, optical and dielectric constants of the material, and in our case, it is also limited by a relatively thin film.

The as-coated CAR films were then baked once at 130 °C for 60 seconds, which is an equivalent of a typical PAB. The baking steps are part of the standard processing of EUV photoresists and are expected to reduce the thickness of the resist significantly.[32,33]

[Si/Mo]$_{60x}$ superlattices with a nominal period of 10 nm were used as a substrate. The first order Bragg reflection in combination with photoelectron spectroscopy yield was used as an independent reference for the incidence angle and also using SW-XPS analysis for depth composition. More details are on using Bragg reflection standing wave spectroscopy and the analysis results can be found in the Supplementary material.

**Theoretical and numerical background**

We will first demonstrate the NTR effects on a model solid/liquid interface. In this model, 20 nm of liquid water is placed on top of bulk SiO$_2$, which is a thick overlayer compared to typical attenuation lengths of photoelectrons. The interfacial region is defined to be 4 nm thick and is equally shared by the solid and the liquid (see Figure 1a). The maximum information depth of the photoemission experiment is still limited by the escape depth of electrons. For the purposes of this thought experiment, we define the information depth to be three times the electron attenuation length, which equals an attenuation factor of ~20 at normal emission. For the example of photelelectrons excited with 3 keV photons, the inelastic mean free path of O 1s core-electrons is around 6 nm in SiO$_2$, resulting in an information depth of approximately 20 nm by our definition.

Using this model interface as an input, the YXRO package[27] (see the Methods section) can compute the electric field profile as a function of depth, the incident angle of the beam. Figure 1b gives one example of such a map for an incident photon energy of 3000 eV. The map serves as a visualization guide to understand the intensity distribution of the X-rays exciting the photoelectrons from different depths when changing the incidence angle. For angles below the critical angle of water (approximately 0.44° using X-ray optical constants from Ref.28), the X-rays start appreciably penetrating the surface layer of water at about 0.2° and generate increasingly strong fields towards the critical angle, which means that the photoelectron spectra originating from the surface layer are the most intense contribution to the overall signal at low angles of incidence. This phenomenon (penetration of X-rays into a medium at total reflection below critical angle) is called evanescent wave and it is a consequence of the boundary conditions for electromagnetic wave in Maxwell equations.[34] With increasing incidence angle, X-rays in the simulation gradually penetrate deeper layers, too, reaching the liquid/solid interface at around incidence of ~0.5°. At this point, the X-rays also enter the solid close to the interface and within a few tenths of one degree, the X-ray penetration depth becomes vastly larger than the escape depth of the photoelectrons.

We now turn our attention to the simulated evolution of the O1s spectral intensity from the four respective regions of the sample as marked in Figure 1a. The dependence of normalized (to the maximum of the photoemission intensity, at so-called Henke[35] peak) intensity on incident angle is shown in Figure 1c and follows the narrative of increasing X-ray penetration depth as displayed in Figure 1b: the signal intensity originating from the topmost layer reaches its maximum at the lowest angle of incidence, while the maximum intensity of deeper layers follows at higher angles of incidence. This means, the incidence angle at which the maximum intensity is reached, is related to the actual depth relative to the surface.

For the purpose of a more detailed discussion, we look at the incidence angle at which the onset of the normalized intensity reaches 10% of its maximum (see Figure 1c). As expected, the topmost 2 nm of liquid ("$H_2O$ surface") exhibits an onset at low incidence angles, close to 0° and reaches 10% at ~0.14°. The "$H_2O$ bulk" region, however, shows an onset at somewhat higher angles, close to 0.2°. Lastly, the intensity of the "interface $H_2O$" photoemission signal shows an onset at significantly higher incidence angles, past 0.4°, followed by a relatively sharp maximum (dashed cyan curve in Fig. 1c). This difference in the signal onset can therefore serve as means to distinguish between surface, bulk and interface contributions. Another striking difference between the "surface", "bulk", and "interface $H_2O$" curves is an oscillatory behavior that the "surface" and "bulk" curves exhibit after reaching their maximum, at around 0.55-0.6° incidence, as highlighted by the vertical dashed lines. This is caused by the presence of the node of the strong standing wave, which suppresses the photoemission signal from the topmost layer of the $H_2O$ film (see white dashed lines in Figure 1b crossing one node and one anti-node of the standing wave).

If we now follow the other two curves that originate from silica (interface and bulk), the angular shift between their offsets is much smaller (~0.01°). But again, the overall shape of the curves, i.e. the width of the peak and intensity decay beyond the maxima, would allow to distinguish them experimentally.

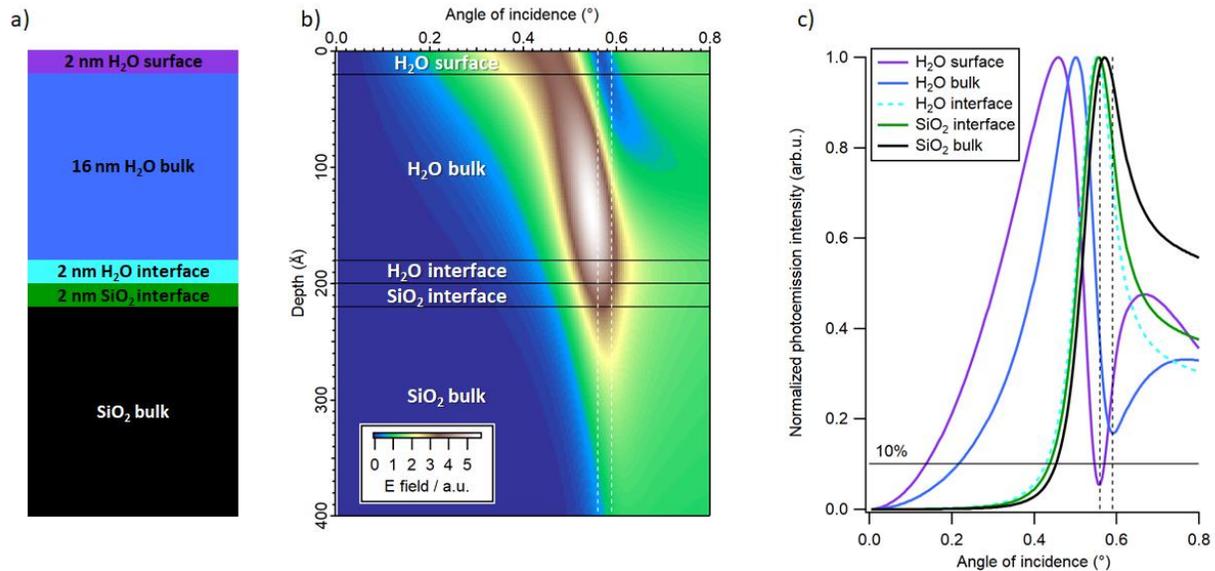

**Figure 1.** a) Schematics of a model solid/liquid system with 2nm thick interface regions. b) Calculated electric field strength as a function of depth and incident angle at excitation energy of 3 keV. c) O1s photoemission intensity as a function of incidence angle for different layers

**Experimental example of 10nm polymer film on Si substrate.**

We now apply the previously described methodology to characterize the interface between the organic photoresist film and the silica/silicon substrate experimentally. These amorphous organic films on the silicon substrate provide many similarities to the solid/liquid interfaces, such as lower optical densities compared to the substrate and a glassy-like phase.

We prepared ~14 nm (nominally) thick films of an organic CAR, which was spin-coated on a Si-terminated [Mo/Si]$_{60x}$ multilayer mirror. Although for the NTR characterization method one does not need a multilayer mirror as a substrate and a normal doped Si substrate is adequate, the Bragg reflection SW-XPS off the multilayer mirror will be used to verify the results of our NTR analysis and to confirm the calibration of the incidence angles. The baking conditions of the sample described herein are part of the standard processing used in the development of EUV photoresists, as shown in Figure 2a.

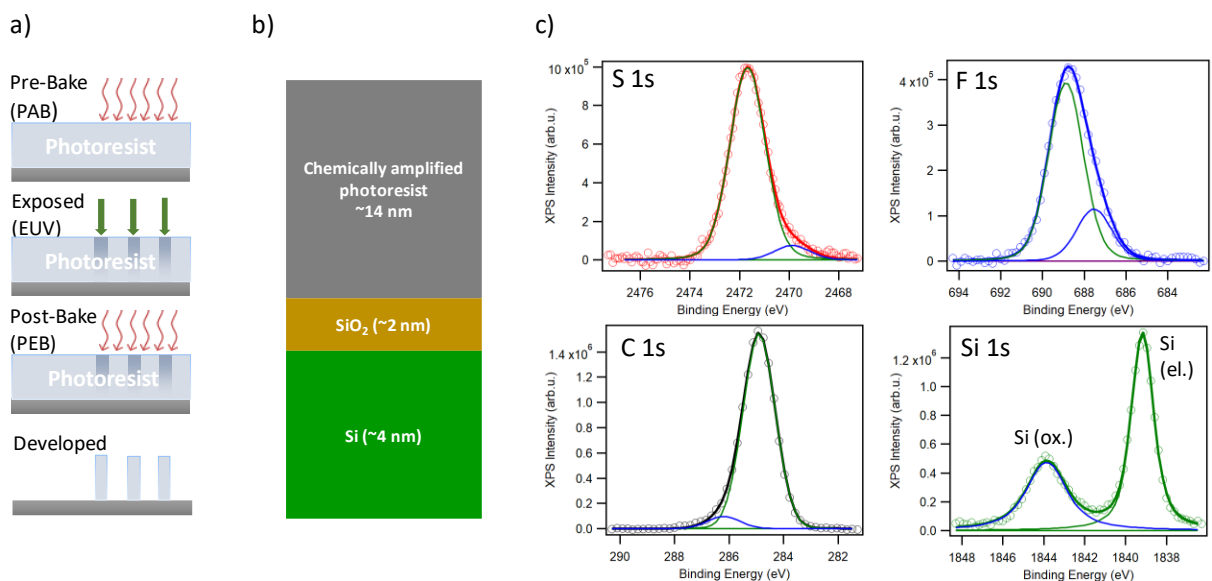

**Figure 2.** a) Schematic of chemically amplified resist (CAR) processing b) Sample schematics including expected/nominal layers' respective thicknesses c) Core levels XPS spectra of the PAB-CAR (S 1s, F 1s, C 1s) and Si 1s from the substrate, exhibiting elemental and oxidized species.

A scheme of PAB sample layers is shown in Figure 2b. The nominal thickness of the photoresist is marked in the drawing. We expect 2 nm of native oxide on the topmost silicon layer of the multilayer mirror (nominally 5nm thick before oxidation). Figure 2c depicts the hard X-ray photoelectron spectra of relevant core-levels measured at the photon excitation energy of 3 keV. A Shirley background was subtracted from all spectra and pseudo-Voigt functions were used to fit individual peaks. Both S 1s and F 1s signals, which originate from the CAR, exhibit two distinctive peak contributions representing two chemical states. The C 1s core-level can also be deconvoluted into two Voigt functions – a low binding energy peak corresponding to C-C and C-H species, and a high binding energy peak corresponding to C=O, C-O, as well as C-F, and C-S species. Si 1s shows two clearly separated peaks - $Si^0$ from the silicon layer and $Si^{4+}$ from its native oxide.

The NTR experiment involves measuring the above core-level spectra at increasing incidence angles, ranging from 0° (parallel to surface) to a few tenths of degrees above critical angle. All measured spectra background-subtracted, and fitted with Voigt functions, as described above. The integrated intensities of relevant photoemission peaks are plotted against the incidence angle in Figure 3a and will be called rocking curves (RCs) in the following.

Two observations are directly apparent from the spectra. Firstly, the order of the low-angle intensity offsets and of the position of the maxima follows the depth of individual elements: Carbon from the photoresist is on top (peak at ~0.55°), followed by silicon oxide (peak at 0.80°) and elemental silicon (peak at 0.84°). Secondly, the curve of silicon oxide exhibits an onset at very shallow incidence (even below Carbon signal), a shoulder at around 0.5° incidence, while its peak intensity is at relatively high incident angle of 0.80°. We will discuss this "anomaly" in the following section.

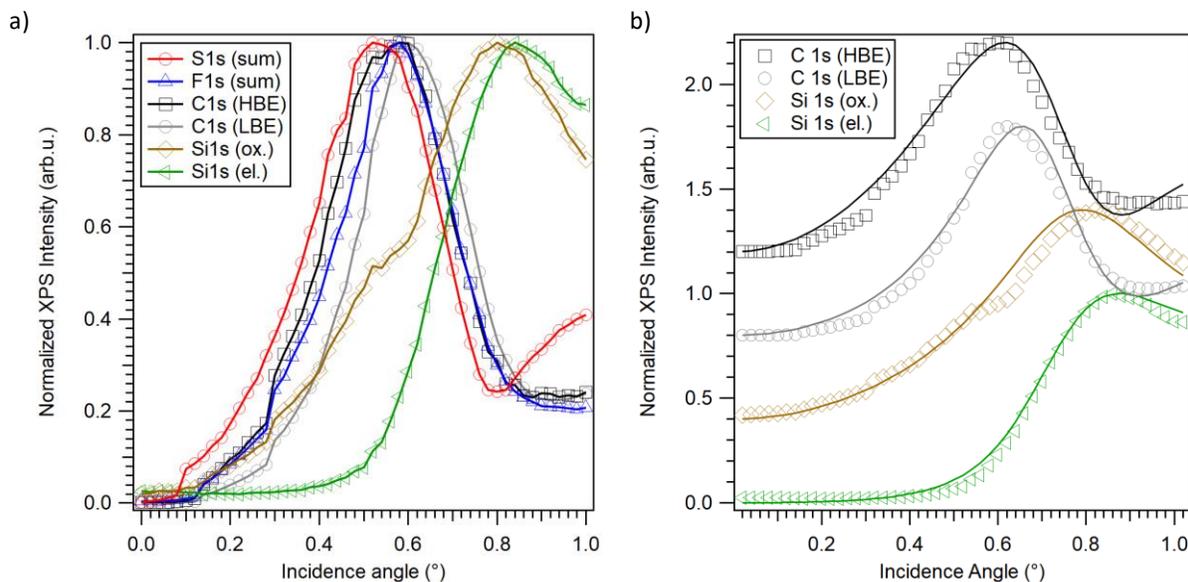

**Figure 3.** a) Experimental rocking curves of relevant core-levels of sulfur, fluorine, carbon (low and high binding energy components) and silicon (elemental and oxide). b) Experimental (markers) and simulated (solid lines) optimized rocking curves for deconvoluted C 1s and Si 1s core-level peaks.

To disentangle the concentration depth-profiles of this system, we compare the experimental data to precise X-ray optical and photoemission simulations. We are using the YXRO simulation package in an iterative way, modifying the model structure to yield best agreement between calculations and experimental data. The process is automated using a global optimization code called SWOPT, described in the Methods section. The optimized structures are shown next to the experimental data in Figure 3b. For the sake of brevity, we are only discussing four rocking curves - two for silicon (elemental and oxidized) and two for carbon (low and high binding energies).

In accordance with the experimental data, the sequence in which simulated rocking curves increase their intensity reflects the order of the species in the sample (from the topmost at lower angles to the most buried layer at higher angles). Qualitatively, the high binding energy carbon (C=O, C-O, C-F, C-S species) is predominantly present in the upper part of the photoresist, while lower binding energy carbon (C-C, C-H species) is spanning all the way to the photoresist/silica interface. In order to get an adequate fit for Si 1s (ox.) peak, the optimizer required a placement of $SiO_x$ contaminant with a sub-monolayer effective thickness on top of the photoresist in our model, which would explain the early onset, the kink at about 0.5°, and the peak maxima at relatively high angle of 0.8° in the experiment and simulation (with the kink being significantly less pronounced in the simulation). We attribute this contamination as a consequence of the sample preparation, including cutting of the silicon wafer, which resulted in some silica residuum on top of our resist film. The structural model and thicknesses corresponding to the optimized simulations are shown in Figures 4a and 4b.

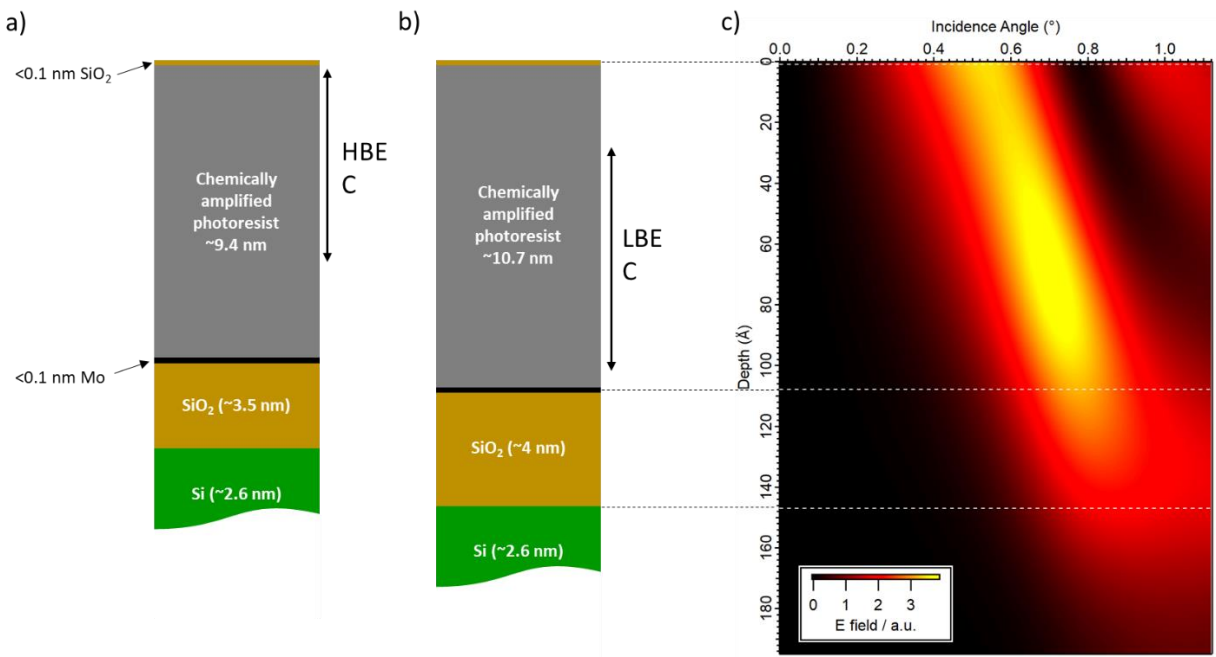

**Figure 4.** a) Concentration depth-profile reflecting the optimized model parameters for LBE Carbon species, and b) HBE Carbon species. b) Electric field strength of the optimized structure as a function of depth and incident angle.

We performed two independent quantitative analyses (for HBE and LBE carbon), with fully relaxed thickness parameters for photoresist, as well as SiOx layers. The optimization process resulted in photoresist thicknesses of 9.4 and 10.7 nm for HBE and LBE datasets (Figure 4a and 4b), respectively, in agreement with the expected thickness of the prepared films (shrunk from nominal value of 14 nm after PAB). A thin (~2.8 nm) topmost layer of the photoresist is formed exclusively by high binding energy carbon species, while the bottom ~2.8 nm of the photoresist consists predominantly low binding energy carbon species. The native silicon oxide layer between the photoresist and the multilayer is with 3.5 nm (HBE carbon analysis) and 4.0 nm (LBE carbon analysis) somewhat thicker than expected for a naturally occurring oxide. The explanation for this phenomenon could be the "wet" deposition of the CAR film, which exposes a surface of silicon to solvents and de-ionized water. The differences in the results of the two analyses also suggest that the accuracy of the method is of the order of ~1 nm, with all the key parameters of the two analyses agreeing within such interval.

The origin of the concentration gradients of the LBE and HBE carbon species at the top and bottom interfaces of the resist can be explained by two phenomena. The top interface, which is directly exposed to air during post-application bake, likely exhibits some degradation and breaking of C-C backbone of the polymer, hence decreasing the concentration of the LBE carbon in the topmost part of the film. On the other hand, the bottom interface is in direct contact with the substrate. It is known, that changes in chemical and physical properties of the photoresists are to a large extent caused by slow secondary electrons, mainly generated by the underlayer.[36] Typical low kinetic energy electron attenuation length in such polymers is ~ 2 – 4 nm,[37] in agreement with our observed thickness of HBE deficient region. Therefore, we expect that the secondary electrons generated during the photoemission experiment caused the changes to the resist chemistries (C=O, C-S, C-F) at the resist/substrate interface.

Another independent analysis using Bragg reflection standing-wave spectroscopy (see Supplementary information) provided a valuable calibration/confirmation of the experimental incidence angles, since it is difficult to experimentally align the beam at true 0° incidence with accuracy better than 0.1°. From the analysis of Mo 3d signal (Supplementary Figure 1a), we discovered a sub-monolayer contaminant layer of Mo at the resist/SiOx interface, which is most likely a remnant of the sputtering process used to grow the superlattice. Therefore, two arrows in the Figure 4a mark the presence of $SiO_x$ and Mo contaminants in sub-monolayer amounts at the two respective interfaces. Even though a conventional XPS depth analysis could be able to detect those impurities, a correct location of these contaminants would be enormously difficult to find without the NTR photoemission data.

On a final note, we want to highlight the simulated electric field strength as a function of incident angle and depth in the sample for a photon energy of 3 keV, which is shown in Figure 4c. Dashed lines serve as guides providing positions of respective interfaces from the sketch in the Figure 4a). An electric field strength cross-section for a particular depth can be used as a first approximation of how the rocking curve of species from that depth would look like. X-rays reach the interface between the resist and silica interface already at an incidence angle of ~0.5°, and reach to the interface of silica and silicon ~0.6°, which agrees with the experimental onset of rocking curves as shown in Figure 3a and b.

Combining these optical effects at near total reflection with ambient pressure photoemission has also its drawbacks. One problem is a relative high analysis complexity. This method requires a monochromatic and angularly well-collimated incident beam, which is available in most synchrotron beamlines, but can be more problematic to achieve in the lab-based systems. Comparing to measuring just at a few angles/excitation energies for a "conventional" depth analysis in XPS, the NTR approach requires rigorous scanning of the incident angle from total reflection past the critical angle in small increments. Another problem is connected to beam spill-over effects. Since the field of view of an ambient pressure hemispherical analyzer is limited by the entrance aperture of the first differentially pumped stage (often less than 1 mm), beam projection on the sample at critical angles (and below) is often much larger than this opening, which leads to a loss of useful signal. The solution to this problem lies in a combination of tightly focused light beam and the elongation of the analyzer aperture in the direction of the longer beam projection. Emerging diffraction limited storage rings of fourth generation synchrotrons are capable of providing (horizontally) highly focused collimated beams that can alleviate this problem. Also, the beam spill-over is much less severe for conventional UHV analyzers, which often have a field of view several millimeters long. The last discussed issue is the absolute calibration of the incident angle. Experimentally, it is very useful to have either a photodiode or a fluorescent screen to observe direct/reflected beam, which allows the determination of the true incident angle. If such instrumentation is not available, the optimization process of the data analysis can compensate small errors in the determination of the incident angle, assuming that densities of the topmost layers are well known. Ref. 15 reports such a NTR-HAXPES study of the perovskite oxide heterostructure, which had no underlying periodic multi-layer and the incident angles were calibrated during the NTR data analysis.

The strengths of the presented NTR method are manifold. It allows for clear disentangling of depth information with very high accuracy (~1 nm) and it does not require simplifications of layered model structures that conventional XPS depth analyses use. The depth scale in this case is "absolute", related to the penetration of X-rays, which is directly connected to the optical density of the material. It is exceptionally powerful in cases when unknown elements occupy different buried depths – as was demonstrated on the case of Mo and $SiO_x$ contaminants in the previous discussion. Lastly, the sample preparation is much more straightforward comparing to Bragg-reflection standing-wave experiments, where multi-layer mirror substrates are necessary as standing-wave generators.

**Conclusions**

We presented a methodology based on the combination of NTR of X-rays and XPS, which can be used for high-precision depth profiling of solid/liquid, solid/solid, and other types of interfaces. We explained this approach on a simulated data for a solid/liquid model interface represented by an $H_2O/SiO_2$ system. We demonstrated the high accuracy of this technique, which can be as high as ~1 nm even for hard X-ray excitation. Secondly, we applied this NTR-XPS to a real system, characterizing an EUV lithography photoresist/silica interface. Here, the disordered, glass-like polymer thin film represents a thin liquid layer like in our simulations and in XPS experiments involving the solid/liquid interface. We were able to identify two different chemical states of carbon in the resist. We determined that the presence of high binding energy carbon attributed to the C=O, C-F, C-S functional groups, is suppressed in the lower part of the photoresist close to resist/silica interface. We attribute this gradient in chemistries at the resist/substrate interface to the effect of secondary electrons generated in the silicon/silica underlayer. On the other hand, the low binding energy carbon attributed to C-C and C-H functional groups has suppressed presence on the top of the film, suggesting the influence of the air exposure after the deposition and during the bake. Using the NTR approach, we were able to measure thicknesses of the respective regions both in the photoresist and in the substrate. These results confirm the high accuracy and interface specificity predicted in our simulations. Accordingly, this method can be applied to study a broad range of interfaces, or any other depth-inhomogeneous system. Its use is especially attractive for solid/liquid interfaces, since it provides all the strengths of a Bragg-reflection standing-wave spectroscopy without the need of an artificial multilayer mirror serving as a standing wave generator.

## Acknowledgements

This research used HAXPES endstation at beamline 9.3.1 of the Advanced Light Source, a U.S. DOE Office of Science User Facility under contract no. DE-AC02-05CH11231. H.P.M. has been supported for salary by the U.S. Department of Energy (DOE) under Contract No. DE-SC0014697.